\documentclass[sigconf]{acmart}

\AtBeginDocument{%
  }

\setcopyright{cc}
\copyrightyear{2026}
\acmYear{2026}
\acmDOI{}
\acmConference[]{CHI 2026 Workshop on Next Steps for Augmented Reality On-the-Move: Challenges \& Opportunities}{Aril 2026}{Barcelona, Spain}
\acmISBN{}

\usepackage{soul}
\usepackage{balance}
\usepackage{natbib}
\usepackage{todonotes}

\copyrightyear{2026}
\acmYear{2026}
\begin{document}

\title{Synchronized Realities: Towards Magic Mobile Experiences through Aligned AR}

\author{Jan Henry Belz}
\orcid{0000-0003-4628-6107}
\affiliation{%
  \institution{Dr. Ing. h.c. F. Porsche AG}
  \city{Stuttgart}
  \country{Germany}
}
\email{jan_henry.belz@porsche.de}

\renewcommand{\shortauthors}{Belz}

\begin{abstract}
  In virtual reality environments, the alignment of perceptual modalities is crucial for immersion and presence. In the AR domain, it is difficult to create such alignments because elements in the physical world are often beyond the user's control. 
  However, recent advances in generative AI enable on-demand content creation, enabling highly reactive AR experiences. Combined with contextual information about the physical world, it has become possible to design experiences that seamlessly align with the user's environment.
  In this reflection paper, I emphasize the importance of "synchronized" realities for context-aware AR experiences, particularly in mobility scenarios. I present several examples of existing synchronized experiences and examine their commonalities and distinctions. Finally, I discuss opportunities and pitfalls of synchronizing AR experiences with the physical world.
\end{abstract}


%   \includegraphics[width=\textwidth]{}
%   \caption{}
%   \Description{}

\maketitle

\section{Introduction}
The alignment of perceptual modalities is a core requirement for immersive virtual reality (VR) scenarios \cite{kim2022studying}. In virtual reality, the entire world can be controlled by the designer or the user; however, this is not possible in augmented reality (AR) scenarios. Physical elements of the real world can only be accessed through embedded hardware, such as in smart home environments (ubiquitous computing) or, in mobility scenarios, in-cabin technology (e.g., cars or buses) \cite{weiser1999computer}. Therefore, the AR experience must adapt to the physical world to provide a seamless, integrated experience \cite{billinghurst2015survey}.

Recent advances in artificial intelligence (AI) research enabled almost human-like zero-shot interpretation of user context (or at least a mimicked version of it) \cite{zhang2024vision}. Natural-language interaction with multimodal perception models enables automated description of a user's environment in "human-readable" ways, e.g., to capture the "vibe" or atmosphere of a moment. Moreover, generative AI allows the creation of customized media content that can be presented to users on demand. For example, information such as images and videos relevant to the current situation can be queried, generated, and displayed in real time. When these methods are combined meaningfully, the user's current situation can be analyzed in depth, and an appropriate user experience can be delivered \cite{hussain2018model}.

While AR generally refers to the visual overlay of information \cite{azuma1997survey}, mixing and augmenting reality with contextual information can occur across multiple modalities (e.g., audio augmented reality (AAR) \cite{bhattacharyya2025birds}). This offers wide opportunities, particularly for mobile applications, such as adaptive music experiences \cite{10.1145/3472749.3474739}. Since mobile scenarios involve rapidly changing contextual information, mobile AR applications need to be even more reactive than in static environments. Therefore, to perfectly align the digital experience with the physical reality, 

The resulting "synchronization" of the digital and physical realities offers enormous potential for seamless, truly ubiquitous user experiences. Through meticulously synchronized experiences, technology can feel like magic, supporting the user in their current activity with zero friction. In this reflection paper, I want to point out the potential of "synchronized reality". I present existing work that proved the effectiveness of such synchronizations, including my own contribution to this area. In three example scenarios, I illustrate commonalities and core aspects of synchronizing AR experiences to the user context. Finally, I discuss opportunities and pitfalls for designing such experiences. 

\section{Example Scenarios}

\paragraph{Synchronized music}
A commuter leaves the office at dusk and starts walking home through a softly lit residential area. Their music player happens to land on a track whose tempo aligns with their pace: the beat sits right on the rhythm of their steps, and the arrangement “opens up” just as they turn onto a quieter street lined with warm streetlights. Nothing in the system explicitly references the sunset or the neighborhood, yet the combination of a steady pace, mellow harmony, and restrained energy makes the moment feel authored, as if the city and the soundtrack are moving in the same meter.

\citet{10.1145/3472749.3474739} create such a meticulous adaptation of the music that the user currently listens to. By synchronizing musical elements with visual landmarks in the environment (such as tunnel exits or highway ramps), they produce an immersive and seamless integration of digital and physical components.

\paragraph{Contextual storytelling}
On a weekend trip, someone drives through a city they don’t know well, listening to an audiobook they selected days earlier. Mid-chapter, the narrator starts describing the very district the listener is currently passing through, mentioning a riverside promenade, a specific bridge silhouette, and the sound of trams in the distance. As the listener looks up and sees the same bridge and hears a tram pass, the story's setting and the physical place snap into alignment.

In my previously published research project \textit{Story-Driven}, I implemented this scenario in a working prototype by utilizing on-demand AI-generated stories. Narrating an audiobook in real time allows for adjusting and guiding the plot in the right direction, no matter where the user is moving \cite{belz2024story}.

\paragraph{Visual AR retrospection}
In the morning, a photo app shows a small "memory" card on the lock screen: a picture from exactly one year ago of friends sitting outside with coffee, wrapped in scarves. Today happens to be similarly cold and bright, and the user is about to meet one of those friends again. The resurfaced image is not just temporally correct; it also matches the day’s atmosphere and upcoming social context, prompting the user to recall the mood, the conversation, and even the route they took.

Similarly, \citet{dingler2016multimedia} captured a continuous stream of images from the user's first-person perspective, which was later used to recall specific moments of the day, thereby augmenting the user's memory using contextual data. Nowadays, wearable AR devices can achieve the same effect on demand, providing an automatic, seamlessly integrated memory aid.

Several other examples exist as research prototypes: stress reduction through an ambient device \cite{maclean2013moodwings} or breathing guide \cite{gemicioglu2024breathe}, real-time adjustment to audience engagement \cite{hassib2017engage}, and adaptive, immersive VR environments \cite{kosunen2016rela} are research areas that have been investigated, built, and evaluated by HCI researchers. While this highlights the potential of meticulous synchronization between the physical and real world, the publication dates also show how sparse the research coverage of such experiences is.

\section{Discussion}
This set of three different scenarios has already been implemented in existing research prototypes. Considering the technical possibilities that context-aware AR combined with generative AI provides, many more scenarios are feasible. When looking at these scenarios from a broader point of view, five key commonalities arise:

\begin{itemize}
    \item \textbf{Anchor:} Every scenario has an initial set of variables that define the situational context. In the example of the synchronized music, this is the visual stimulus of the environment ("softly lit residential area") and the rhythm of the user's steps. Identifying and sensing the anchor is crucial for providing a well-synchronized experience.
    \item \textbf{Precision:} Depending on the type of user experience, elements may have to be adjusted more or less precisely, from multiple times per second to only once per hour. For example, in our storytelling project \textit{Story-Driven}, we checked for new data every five seconds \cite{belz2024story}, whereas adaptive music playlisting only requires synchronization every couple of minutes \cite{wang2012context}.
    \item \textbf{Trigger:} Similar to the precision of adjustment, the type of trigger depends on the scenario: Sometimes, experience \textbf{boundaries} trigger a new experience cycle (e.g., a visual landmark appears \cite{10.1145/3472749.3474739}), in other cases, the adaptation process is a \textbf{continuous} one (e.g., \textit{Story-Driven} \cite{belz2024story}).
    \item \textbf{Interaction type:} Experiences can be synchronized \textbf{implicitly} based solely on contextual data. However, it can be important to require \textbf{explicit} user input before adjusting the current user experience.
    \item \textbf{Stakes:} Some adjustments to a user experience can impose risks to the user, for example, through distraction or invasion of privacy. Particularly failure of synchronization can cause discomfort. Therefore, synchronized experiences require careful consideration of their relevance to the user and may be confirmed manually by the user (see \textit{explicit interaction}).
\end{itemize}

In \textit{Story-Driven}, we implemented a synchronized storytelling experience that adjusts to the user's current location. For example, if the user is walking towards a historic building or landmark, that place becomes the setting for the upcoming plot. Moreover, the story is synchronized at a fine level, so the place name is mentioned precisely when the user arrives \cite{belz2024story}. In this case, the \textbf{anchor} is the point of interest that the user is heading towards, as well as the user's estimated time of arrival. The experience \textbf{precision} is medium-fine, since the story is adjusted every five seconds, but the time window for the user arriving at the landmark leaves room for error. Synchronization \textbf{triggers} in \textit{Story-Driven} are continuous, since the user's estimated time of arrival is checked every five seconds, which triggers a story update. The \textbf{interaction type} is purely implicit; No user input is required. Finally, the \textbf{stakes} are quite low: if the synchronization fails, the story is still narrated, but at a different setting - no harm done to the user besides them being slightly bored.

Besides the main outcome of the user study, the design process of our research project revealed interesting insights for synchronized experiences. First, the anchor needs to be chosen carefully, as it determines the effectiveness of a synchronized experience. The mobile experience of \textit{Story-Driven} would not have been transferable to other modalities if we had selected something other than the user's estimated time of arrival (e.g., speed would be significantly different for a car than for pedestrians). Second, the synchronization precision must be defined appropriately. Querying and processing contextual information every second might not be feasible in mobile scenarios with low computational power. Updating the context only every ten seconds might leave too much room for error, resulting in an out-of-sync experience. A suitable selection of boundary-based triggers combined with a feasible synchronization precision can result in a better experience than simple high-frequency update cycles. Finally, meticulously synchronized experiences are prone to fail and run out of sync - experience designers need to consider this and prepare for failure in consideration of the stakes.

Synchronized realities offer a large potential for immersive, frictionless, and "calm" experiences, both in visual and multimodal AR. It may take additional efforts to create a robust user experience that accounts for various contexts, but it also opens up the possibility for an almost "magical" experience. Context-aware, mobile AR offers the richest potential for magical experiences. In this reflection paper, I want to emphasize the importance of well-synchronized user experiences that "weave themselves into the fabric of everyday life until they are indistinguishable from it" \cite{weiser1999computer}. 

\section{Conclusion}
% What other scenarios could be implemented? What "ground rules" can we abstract from these scenarios for fine-grained synchronization?
% Is there a group of scenarios that could be implemented by one dynamic system?

% \begin{acks}

% \end{acks}

\bibliographystyle{ACM-Reference-Format}
\bibliography{references}

\end{document}